\documentclass[10pt,conference,letterpaper]{IEEEtran}
\usepackage{epsfig}
\usepackage{epstopdf}
\usepackage{times}
\usepackage{algorithm}
\usepackage{algorithmic}
\usepackage{amssymb}
\usepackage{amsmath}
\usepackage{subfigure}
\usepackage{multirow}
\usepackage{url}
\usepackage{eso-pic}
\usepackage{lipsum}
\usepackage{soul}
\usepackage{caption}
\usepackage{mathtools}
\usepackage{bbm}
\usepackage{framed}

\usepackage{array} 
\def\ignore#1\endignore{}
\newcolumntype{h}{@{}>{\ignore}l<{\endignore}} 

\newcolumntype{x}[1]{%
>{\centering\hspace{0pt}}p{#1}}%

\usepackage{latexsym}
\usepackage{multicol}
\usepackage{eurosym}
\usepackage{amsthm}
\usepackage{xspace}
\usepackage{pdfpages}
\usepackage{verbatim}
\usepackage{stfloats}

\newenvironment{sketch}{{\noindent \it Sketch of Proof:~}}

\newtheorem{proposition}{Proposition}

\newtheorem{lemma}{Lemma}

\newcommand{\norm}[1]{\left\lVert#1\right\rVert}

\newcommand{\vv}[1]{\boldsymbol{#1}}

\renewcommand{\baselinestretch}{0.965}

\begin{document}
\title{STORNS: Stochastic Radio Access Network Slicing}

\author{
Vincenzo Sciancalepore\IEEEauthorrefmark{1},
Marco Di Renzo\IEEEauthorrefmark{2},
Xavier Costa-Perez\IEEEauthorrefmark{1}\\
\IEEEauthorrefmark{1} NEC Laboratories Europe GmbH\quad\IEEEauthorrefmark{2} CNRS -
CentraleSupelec - Univ. Paris-Sud
}
\maketitle
\psfull

\setlength{\textfloatsep}{1pt}

\begin{abstract}
Recently released 5G networks empower the novel \emph{Network Slicing} concept. 
Network slicing introduces new business models such as allowing telecom providers to lease a virtualized slice of their infrastructure to \emph{tenants} such as industry verticals, e.g. automotive, e-health, factories, etc. However, this new paradigm poses a major challenge when applied to Radio Access Networks (RAN): \emph{how to achieve revenue maximization while meeting the diverse service level agreements (SLAs) requested by the infrastructure tenants?}

In this paper, we propose a new analytical framework, based on stochastic geometry theory, to model realistic RANs that leverage the business opportunities offered by network slicing. 
We mathematically prove the benefits of slicing radio access networks as compared to non-sliced infrastructures. Based on this, we design a new admission control functional block, STORNS, which takes decisions considering per slice SLA guaranteed average experienced throughput. A radio resource allocation strategy is introduced to optimally allocate transmit power and bandwidth (i.e., a slice of radio access resources) to the users of each infrastructure tenant. Numerical results are illustrated to validate our proposed solution in terms of potential spectral efficiency, and compare it against a non-slicing benchmark.
\end{abstract}

\section{Introduction}
\label{s:intro}
Upcoming service requirements from vertical industries call for a novel design of mobile networks, namely 5G. Key-enablers have been identified as network programambility and virtualization:
the former brings the benefits of automation and reactiveness of software modules, allowing to (re)configure mobile networks dynamically while in operation; the latter overcomes the limitations of monolithic network infrastructures by abstracting the concept of ``network function'' and providing flexibility in composing, placing and managing these functions. In this context, the novel definition of \emph{network slicing}~\cite{slicing_conceptNGMN} encompasses such new requirements and constitutes an enabler for potential economical benefits. New vertical industries, e.g., automotive, e-health, factories etc., are entering into the telecom market and are disrupting the traditional business models of telecom operators. They are forcing infrastructure providers to open their networks to tenants, a solution that provides incentives for monetizing the availability of isolated and secure (virtualized) network slices~\cite{HSFCS_INFOCOM2019,DemoSlicing2016}.

\begin{figure}[t!]
      \centering
      \includegraphics[trim = 0mm 0mm 0mm 0mm, clip, width=0.7\linewidth ]{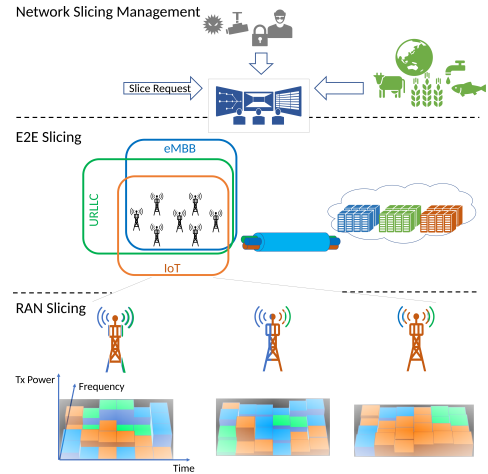}
      \caption{Illustration of the Network Slicing Concept}
      \label{fig:arch}
\end{figure}

This new disruptive concept has spurred research interest in both academic and industrial communities. Its realization, however, requires the solution of a number of technical challenges that, for the time being, are not completely resolved~\cite{NetworkSlicing_Survey}. In the future, telecom providers envision an increasing demand for end-to-end network slices, which involve heterogeneous service level agreements (SLAs) comprising different key performance indicators (KPIs), such as throughput, latency and reliability~\cite{SSC_INFOCOM2017}. However, this requires appropriate automated admission control and resource allocation protocols for designing efficient network management systems~\cite{NEC_Conext2018}. In particular, the resource management of the radio access network (RAN) is one of the most challenging aspects that needs to be dealt with. In~\cite{NetworkSharing2016SCS}, a first network slicing brokering solution has been presented for automated network slicing admission control decisions. We build on that architecture to illustrate in Fig.~\ref{fig:arch} some of the challenges that need to be solved for efficiently slicing the RAN. The available resources of the air interface can be sliced at multiple levels: in frequency, time, and power domains. Slicing at this level of granularity requires to account for the cellular network topology, the other-cell interference and the radio channel conditions experienced by the users of every single tenant~\cite{TC2018_slicing}. To solve this never-addressed and challenging issue, we leverage the mathematical tool of stochastic geometry and the theory of point processes~\cite{tutorial_stoch_2017}. 
To the best of our knowledge, we pioneer the design of an automated RAN slicing admission control and resource allocation scheme that provides throughput guarantees in the RAN, where the cellular network topology and the other-cell interference are taken into account.

Our main research contributions can be summarized as follows: $i$) we derive a new formulation of the network spectral efficiency with the aid of stochastic geometry tools explicitly accounting for the interplay among the transmit power of the cellular base stations (BSs), the available spectrum and the deployment densities of cellular BSs and mobile terminals (MTs) of each tenant, $ii$) we study the analytical properties of the newly proposed utility function for slicing the RAN and mathematically prove its convexity, and $iii$) we design a RAN admission control and a novel STOchastic RaN Slicing mechanisms (\emph{STORNS}). STORNS accounts for slice SLAs in terms of average throughput and assigns, in an automated fashion, time/frequency resources and transmit power levels to each slice of the tenants admitted into the sliced RAN.

\section{System Model} \label{s:system}
We explicitly account for the topology of cellular networks by using the mathematical tools of stochastic geometry and point processes~\cite{ABG_TC11}. Under a stochastic geometry framework, in particular, the locations of BSs and MTs are modeled as points of a point process with some specific spatial properties, due to its tractability.

We consider a RAN with multiple access points (i.e., the BSs), such as long-term evolution (LTE) eNBs, femto-cells, mm-wave access points. Multiple tenants are available in the network. The generic infrastructure tenant, $i\in\mathcal{I}$, is willing to pay for managing a ``slice'' of the resources of the RAN, provided that a certain slice SLA is guaranteed to it, e.g., a minimum average throughput requirement. Users $u\in\mathcal{U}_i$ (i.e., the MTs) belong to a particular tenant $i$ and are assumed to have density $\lambda_{\rm{Ti}}$. We consider a single network operator making available its physical resources to the multiple tenants. In particular, each tenant accesses the RAN resources so that its users share common resources of the (same) air interface with specific privileges. Specifically, the BSs of the cellular network operate in an open access mode for all the users of each tenant. However, each BS serves different tenants in a non-overlapping frequency band and by using part of its available total transmit power. As a consequence, each tenant owns a dedicated part of the spectrum and of the transmit power, which constitute the ``isolated'' slice of the physical resources requested to the network operator. It is worth noting that the requests of the tenants are not related to the specific spectrum share or transmit power that are eventually assigned to them by the network operator. The tenants are, on the other hand, interested in getting a minimum required spectral efficiency (expressed in bit/sec/m$^2$), which allows them to satisfy the specific service requirements of their own users, regardless of the presence of the other tenants in the network. In this paper, we formulate the minimum spectral efficiency requested by each tenant as a percentage of the spectral efficiency of the network without slicing the RAN, i.e., when tenants do not request any guaranteed service to the network operator. Based on these assumptions, we formulate an optimization problem and identify the optimal transmit power and spectrum to be assigned to each tenant so as to obtain the requested spectral efficiency. The solution of this system-level optimization problem provides insights on the advantages of a sliced network, and sheds light on the feasible set of spectral efficiencies that each tenant can request as a function of the network throughput without applying network slicing. In the sequel, for ease of description, the system model is introduced by considering a two-tenant scenario. The two tenants are denoted by T$1$ and T$2$. The generalization to more than two tenants is detailed in Section~\ref{sect:gen_problem}.

%
\subsection{Cellular Network Modeling} \label{PPP_CellularModeling}
The BSs are modeled as points of a homogeneous Poisson point process (PPP), denoted by $\Psi_{\rm{BS}}$, of density $\lambda_{\rm{BS}}$.
The MTs of each tenant are modeled following a different homogeneous PPP, denoted by $\Psi_{\rm{Ti}}$, of density $\lambda_{\rm{Ti}}$ for $i=1,2$. $\Psi_{\rm{BS}}$, $\Psi_{\rm{T1}}$ and $\Psi_{\rm{T2}}$ are assumed to be independent. The MTs are served by the BS providing the best average received power on the downlink channel. All the other BSs transmitting over the same frequency spectrum act as interfering BSs (i.e., full-frequency reuse is considered). Each BS transmits with constant power. $P_{\rm{tot}}$ denotes the total power budget of each BS. Each BS transmits in a spectrum of total bandwidth $B_{\rm{tot}}$. The percentage of transmit power and bandwidth used by ${\rm{Ti}}$ are denoted by $P_{\rm{Ti}}$ and $B_{\rm{Ti}}$ for $i=1,2$, respectively, such that $P_{\rm{T1}}+P_{\rm{T2}} \le P_{\rm{tot}}$ and $B_{\rm{T1}}+B_{\rm{T2}} \le B_{\rm{tot}}$. The spectrum bands used by $\rm{T1}$ and $\rm{T2}$ are non-overlapping and, thus, no inter-tenant interference is available. 

Let us consider a generic BS of the network. All the MTs of $\rm{T1}$ and $\rm{T2}$ served by this BS equally share the available transmit power and bandwidth, i.e., power and spectrum are viewed as continuous resources by the BS's scheduler and, thus, no intra-cell interference is available\,\footnote{Considering continuous resources (e.g., bandwidth) makes our analysis tractable and easy to explain. However, this assumption can be relaxed by accounting for discrete resources (e.g., physical resource blocks (PRBs))~\cite{RLG_TWC16}.}. As a result, the transmit power spectral density of the BSs of tenant ${\rm{Ti}}$ is ${{{P_{{\rm{Ti}}}}} \mathord{\left/ {\vphantom {{{P_{{\rm{Ti}}}}} {{B_{{\rm{Ti}}}}}}} \right. \kern-\nulldelimiterspace} {{B_{{\rm{Ti}}}}}}$ for $i=1,2$. This implies that a BS is off only if there are no MTs, either from $\rm{T1}$ or $\rm{T2}$, within its corresponding coverage region. If ${\rm{Ni}}$ MTs belong to tenant ${\rm{Ti}}$ for $i=1,2$, this implies that each MT uses bandwidth ${{{B_{{\rm{Ti}}}}} \mathord{\left/{\vphantom {{{B_{{\rm{Ti}}}}} {{N_{{\rm{Ti}}}}}}} \right.\kern-\nulldelimiterspace} {{N_{{\rm{Ti}}}}}}$ and that the MTs do not interfere with each other. The other-cell interference (among BSs of the same tenant transmitting over the same spectrum) is, on the other hand, taken into account.

Based on this system model, in the next section we formulate the potential spectral efficiency (PSE), i.e., the average network throughput, in bit/sec/m$^2$ for each tenant of the network, by either using or not network slicing. In the latter case, the tenants equally share the resources of the network operator without any constraints on their minimum service requirements. In this case, in other words, tenants $\rm{T1}$ and $\rm{T2}$ equally share the transmit power $P_{\rm{tot}}$ and bandwidth $B_{\rm{tot}}$.
\subsection{Potential Spectral Efficiency} \label{PSE}
For ease of notation, we formulate the PSE for a generic tenant whose MTs constitute a PPP of density $\lambda_{\rm{T}}$ and whose BSs allocate transmit power $P$ and bandwidth $B$. The PSE can be formulated as follows:
\begin{equation}
\label{Eq_1}
{\rm{PSE}}\left( {P,B,{\lambda _{\rm{T}}}} \right) = \sum\limits_{n = 0}^{ + \infty } {{\rm{PSE}}\left( {\left. {P,B,{\lambda _{\rm{T}}}} \right|n + 1} \right){P_r}\left\{ n,{\lambda _{\rm{T}}} \right\}}
\end{equation}
\noindent where ${{\rm{PSE}}\left( {\left. {P,B,{\lambda _{\rm{T}}}} \right|n + 1} \right)}$ is the PSE by conditioning on the number, $n+1$, of MTs in a generic cell and ${{P_r}\left\{ n,{\lambda _{\rm{T}}} \right\}}$ is the probability that, given a MT in a cell, there are other $n$ MTs in it.

Let $\gamma_I$ be the reliability threshold for successfully decoding a data packet and $\gamma_A$ be the reliability threshold for detecting the presence of the serving BS during the cell association phase. With the aid of stochastic geometry (\cite{ABG_TC11} and \cite{RLG_TWC16}), ${{\rm{PSE}}\left( {\left. {P,B,{\lambda _{\rm{T}}}} \right|n + 1} \right)}$ and ${{P_r}\left\{ n,{\lambda _{\rm{T}}} \right\}}$ can be formulated as follows:
\begin{equation}
\label{Eq_2}
\begin{split}
{\rm{PSE}}\left( {\left. {P,B,{\lambda _{\rm{T}}}} \right|n + 1} \right) &= {\lambda _{\rm{T}}}\frac{B}{{n + 1}}{\log _2}\left( {1 + {\gamma _I}} \right)\\
 & \hspace{-1.5cm} \times {P_r}\left\{ {{\rm{SIR}}\left( {n + 1} \right) \ge {\gamma _I},\overline {{\rm{SNR}}} \left( {n + 1} \right) \ge {\gamma _A}} \right\}
\end{split}
\end{equation}
\begin{equation}
\label{Eq_3}
{P_r}\left\{ n,{\lambda _{\rm{T}}} \right\} = \frac{{{{3.5}^{3.5}}\Gamma \left( {n + 4.5} \right){{\left( {{{{\lambda _{\rm{T}}}} \mathord{\left/
 {\vphantom {{{\lambda _{\rm{T}}}} {{\lambda _{{\rm{BS}}}}}}} \right.
 \kern-\nulldelimiterspace} {{\lambda _{{\rm{BS}}}}}}} \right)}^n}}}{{\Gamma \left( {3.5} \right)\Gamma \left( {n + 1} \right){{\left( {3.5 + {{{\lambda _{\rm{T}}}} \mathord{\left/
 {\vphantom {{{\lambda _{\rm{T}}}} {{\lambda _{{\rm{BS}}}}}}} \right.
 \kern-\nulldelimiterspace} {{\lambda _{{\rm{BS}}}}}}} \right)}^{n + 4.5}}}}
\end{equation}
\noindent where $\Gamma(\cdot)$ is the gamma function, ${{\rm{SIR}}\left( {n + 1} \right)}$ and ${\overline {{\rm{SNR}}} \left( {n + 1} \right)}$ are the signal-to-interference-ratio (SIR) and the average signal-to-noise-ratio (SNR), given the number of MTs, $n+1$, in a generic cell, during the information decoding and the cell association phases, respectively. They are defined as follows:
\begin{equation}
\label{Eq_4}
{\rm{SIR}}\left( {n + 1} \right) = \frac{{{{\left( {{P \mathord{\left/
 {\vphantom {P {(n + 1)}}} \right.
 \kern-\nulldelimiterspace} {(n + 1)}}} \right){h_0}} \mathord{\left/
 {\vphantom {{\left( {{P \mathord{\left/
 {\vphantom {P {(n + 1)}}} \right.
 \kern-\nulldelimiterspace} {(n + 1)}}} \right){h_0}} {{L_0}}}} \right.
 \kern-\nulldelimiterspace} {{L_0}}}}}{{\sum\limits_{k \in {\Psi _{{\rm{BS}}}}} {{{\left( {{P \mathord{\left/
 {\vphantom {P {(n + 1)}}} \right.
 \kern-\nulldelimiterspace} {(n + 1)}}} \right){h_k}} \mathord{\left/
 {\vphantom {{\left( {{P \mathord{\left/
 {\vphantom {P {(n + 1)}}} \right.
 \kern-\nulldelimiterspace} {(n + 1)}}} \right){h_k}} {{L_k}}}} \right.
 \kern-\nulldelimiterspace} {{L_k}}}{\bf{1}}_{\left( {{L_k} > {L_0}} \right)}} }}
\end{equation}
\begin{equation}
\label{Eq_5}
\overline {{\rm{SNR}}} \left( {(n + 1)} \right) = \frac{{{{\left( {{P \mathord{\left/
 {\vphantom {P {(n + 1)}}} \right.
 \kern-\nulldelimiterspace} {(n + 1)}}} \right)} \mathord{\left/
 {\vphantom {{\left( {{P \mathord{\left/
 {\vphantom {P {(n + 1)}}} \right.
 \kern-\nulldelimiterspace} {(n + 1)}}} \right)} {{L_0}}}} \right.
 \kern-\nulldelimiterspace} {{L_0}}}}}{{{N_0}\left( {{B \mathord{\left/
 {\vphantom {B {(n + 1)}}} \right.
 \kern-\nulldelimiterspace} {(n + 1)}}} \right)}}
\end{equation}
\noindent where ${{h_0}}$ and ${{h_k}}$ are the fading power gains of the serving and interfering BSs of a generic MT, respectively, due to the wireless channels, which are assumed to be independent and identically distributed exponential random variables with unit mean, ${L_0} = \kappa r_0^\beta$ and ${L_k} = \kappa r_k^\beta$ are the path-losses of serving and interfering BSs, respectively, where $\kappa$ is the path-loss propagation constant, $\beta > 2$ is the path-loss exponent, and $r_0$ and $r_k$ are the distances of serving and interfering BSs, and $N_0$ is the noise-power spectral density. The indicator function ${{\bf{1}}_{\left( {{L_k} > {L_0}} \right)}}$ accounts for the cell association strategy and implies that the path-loss of the serving BS, averaged over the fast fading, is smaller than the path-losses of the interfering BSs. It is worth mentioning that $\overline {{\rm{SNR}}} \left( {n + 1} \right)$ is averaged with respect to the fast fading in order to avoid frequent handovers due to the channel variations, as usual in cellular networks.

By comparing \eqref{Eq_2} with the typical definition of PSE in \cite{ABG_TC11} and \cite{RLG_TWC16}, we note that our definition is more realistic since it accounts for non-zero values of $\gamma_{A}$ and, thus, for the fact that the MTs cannot detect an arbitrary weak signal. This is a fundamental change of our modeling, which allows us to obtain an expression of the PSE that explicitly depends on the transmit power of the BSs. The PSE is reported in the following proposition.
\begin{proposition}
The exact mathematical expression of the PSE is given in Eq.~\eqref{Eq_6} below:
\end{proposition}
%
\vspace{-10mm}
\begin{equation}
\label{Eq_6}
\begin{split}
& {\rm{PSE}}\left( {P,B,{\lambda _{\rm{T}}}} \right) = B{\log _2}\left( {1 + {\gamma _I}} \right)\frac{{{\lambda _{{\rm{BS}}}}L\left( {\frac{{{\lambda _{\rm{T}}}}}{{{\lambda _{{\rm{BS}}}}}}} \right)}}{{1 + L\left( {\frac{{{\lambda _{\rm{T}}}}}{{{\lambda _{{\rm{BS}}}}}}} \right)\Upsilon \left( {{\gamma _I},\beta } \right)}}\\
& \times \left[ {1 - \exp\!\!\left( {\!\! -\pi {\lambda _{{\rm{BS}}}}{{\left( {{\tau _A}\frac{P}{B}} \right)}^{{2 \mathord{\left/
 {\vphantom {2 \beta }} \right.
 \kern-\nulldelimiterspace} \beta }}}\!\!\!\left(\! {1 + L\!\left( {\frac{{{\lambda _{\rm{T}}}}}{{{\lambda _{{\rm{BS}}}}}}} \right)\Upsilon \left( {{\gamma _I},\beta } \right)} \right)} \right)} \right]
\end{split}
\end{equation}
\noindent where ${\tau _A} = {\left( {\kappa {\gamma _A}{N_0}} \right)^{ - 1}}$ and:
\begin{equation}
\label{Eq_7}
L\left( {\frac{{{\lambda _{\rm{T}}}}}{{{\lambda _{{\rm{BS}}}}}}} \right) = 1 - {\left( {1 + \frac{1}{{3.5}}\frac{{{\lambda _{\rm{T}}}}}{{{\lambda _{{\rm{BS}}}}}}} \right)^{ - 3.5}} \ge 0
\end{equation}
\begin{equation}
\label{Eq_8}
\Upsilon \left( {{\gamma _I},\beta } \right) = {}_2{F_1}\left( { - \frac{2}{\beta },1,1 - \frac{2}{\beta }, - {\gamma _I}} \right) - 1 \ge 0
\end{equation}
\noindent with ${}_2{F_1}\left(  \cdot  \right)$ denoting the Gauss hypergeometric function.
\begin{sketch}
Eq.~\eqref{Eq_6} is obtained from the definition of PSE in~\cite[Eq. (15)]{RLG_TWC16}, by computing the coverage probability with the aid of mathematical steps similar to those in~\cite[Theorem 1]{ABG_TC11}. The difference with respect to~\cite[Theorem 1]{ABG_TC11} lies in the non-zero value of $\gamma_A$, which modifies the upper-limit in~\cite[Eq. (2)]{ABG_TC11} from $L_0 \to \infty$ to $\frac{1}{L_0} \le N_0 B \gamma_A / P$. The proof follows by taking into account the definition of $L_0$, i.e., $L_0 = \kappa r_0^\beta$, and by solving the integral in closed-form.
\qed
\end{sketch}

If $\gamma_{A} = 0$, it is worth nothing that the PSE is independent of the transmit-power, $P$, of the BSs and the proposed framework simplifies to previously reported formulas in \cite{ABG_TC11} and \cite{RLG_TWC16}. In addition, the PSE would linearly depend on the transmission bandwidth $B$. By considering that the BSs have a finite sensitivity for detecting the presence of the BSs, i.e., $\gamma_{A} \ne 0$, on the other hand, we obtain a more accurate mathematical framework where $P$ and $B$ play a fundamental role for system optimization in the context of a multi-tenant cellular network with network slicing capabilities.

\section{System-Level Optimization} \label{OPT}
Based on the mathematical formulation of the PSE in \eqref{Eq_6}, the PSE of tenant ${\rm{Ti}}$ in the presence of network slicing is ${\rm{PS}}{{\rm{E}}_{{\rm{Ti}}}} = {\rm{PSE}}\left( {{P_{{\rm{Ti}}}},{B_{{\rm{Ti}}}},{{\lambda _{{\rm{Ti}}}}}} \right)$ for $i=1,2$. Conversely, the PSE without a network slicing---MTs of $\rm{T1}$ and $\rm{T2}$ share the available resources of the RAN without any service requirement constraints---is ${\rm{PS}}{{\rm{E}}_{{\rm{NoSlicing}}}} = {\rm{PSE}}\left( {{P_{{\rm{tot}}}},{B_{{\rm{tot}}}},{\lambda _{{\rm{tot}}}} = {\lambda _{{\rm{T1}}}} + {\lambda _{{\rm{T2}}}}} \right)$.

%
%
%
Let us now consider a toy scenario. 
We assume a homogeneous network deployment, e.g., all the BSs have the same bandwidth $B_{\rm{tot}}$ and maximum transmission power $P_{\rm{tot}}$. The tenant SLAs are formulated in terms of average PSE\,\footnote{The PSE is interpreted as the average network throughput experienced by the users of the tenant. This is a reasonable assumption when considering tenant SLAs in terms of cell throughput.}, i.e., ${\rm{PS}}{{\rm{E}}_{{\rm{Ti}}}} = \alpha_{{\rm{Ti}}}{\rm{PS}}{{\rm{E}}_{{\rm{NoSlicing}}}}$ where ${\alpha _{{\rm{Ti}}}} \ge 0, \forall i$, which constitute the minimum spectral efficiency requirements of tenant $\rm{T1}$ and $\rm{T2}$, respectively. The SLAs, in particular, are expressed as a fraction of the spectral efficiency without performing network slicing, i.e., the baseline working operation of current cellular networks.

Let us introduce the short-hand notation: $k_1^{(\rm{Ti})} = {\log _2}\left( {1 + {\gamma _I}} \right)\frac{{{\lambda _{{\rm{BS}}}}L\left( {\frac{{{\lambda _{\rm{Ti}}}}}{{{\lambda _{{\rm{BS}}}}}}} \right)}}{{1 + L\left( {\frac{{{\lambda _{\rm{Ti}}}}}{{{\lambda _{{\rm{BS}}}}}}} \right)\Upsilon \left( {{\gamma _I},\beta } \right)}}$ and $k_2^{(\rm{Ti})} = \pi {\lambda _{{\rm{BS}}}}{{\left( {{\tau _A}} \right)}^{{2 \mathord{\left/ {\vphantom {2 \beta }} \right. \kern-\nulldelimiterspace} \beta }}}\!\!\!\left(\! {1 + L\!\left( {\frac{{{\lambda _{\rm{Ti}}}}}{{{\lambda _{{\rm{BS}}}}}}} \right)\Upsilon \left( {{\gamma _I},\beta } \right)} \right)$. The following optimization problem can be formulated.

\vspace{2mm}\noindent \textbf{Problem}~\texttt{Bi-Sharing}:
\begin{equation*}
\label{pr:bi-sharing}
\begin{array}{ll}
\text{minimize\!\!\!}  & \mathbbm{1}\\
\text{subject to\!\!\!} & k_1^{(\rm{T1})}\!B_{\rm{T1}}\!\left(1\! -\! e^{-(\frac{P_{\rm{T1}}}{B_{\rm{T1}}})^{(2/ \beta)}k_2^{(\rm{T1})}}\!\right)\!\! \geq\!\alpha_{\rm T1} {\rm{PS}}{{\rm{E}}_{{\rm{NoSlicing}}}};\\
				   & k_1^{(\rm{T2})}\!B_{\rm{T2}}\!\left(1\! -\! e^{-(\frac{P_{\rm{T2}}}{B_{\rm{T2}}})^{(2/ \beta)}k_2^{(\rm{T2})}}\!\right)\!\! \geq\!\alpha_{\rm T2} {\rm{PS}}{{\rm{E}}_{{\rm{NoSlicing}}}};\\
				   & B_{\rm T1} + B_{\rm T2} \leq B_{\rm{tot}};\\
				   & P_{\rm T1} + P_{\rm T2} \leq P_{\rm{tot}};\\
                   & B_{\rm T1}, B_{\rm T2}, P_{\rm T1}, P_{\rm T2} \in \mathbb{R}_{+};
\end{array}
\end{equation*}
where the total throughput without slicing is ${\rm{PS}}{{\rm{E}}_{{\rm{NoSlicing}}}} = k_1^{(\rm{tot})}B_{\rm{tot}}\!\left( 1\!-\!e^{-(\frac{P_{\rm{tot}}}{B_{\rm{tot}}})^{(2/ \beta)}k_2^{(\rm{tot})}} \right)$.
%
%
Generally speaking, Problem~\texttt{Bi-Sharing} provides the optimal set of values $\vv{b} = \{ B_{\rm i}\}$ and $\vv{p} = \{ P_{\rm i}\}$ given the tenants SLAs.

\subsection{The relevance of a sliced RAN}

The previous example provides the baseline scenario for our analysis. It unveils important insights when applying the network slicing concept to the RAN of cellular networks. 

\begin{lemma}\label{lemma:unsliced}
The probability that the sum-PSE experienced by all the tenants is greater than the PSE experienced by a monolithic non-sliced network is greater than zero, i.e., $\text{Pr}\{ (\rm PSE_{\rm{T1}}+ \rm PSE_{\rm{T2}})\geq \rm {PSE}_{{NoSlicing}}\}\geq \epsilon, \forall \epsilon>0$.
\end{lemma}
This lemma relies on the convexity property of the multivariable PSE function shown in Eq.~\eqref{Eq_6}\,\footnote{Note that the Hessian condition for that function is not fully satisfied. Therefore, as shown in~\cite{Boyd2004}, it is needed to check that $g(tx_1 + (1-t)x_2,ty_1 + (1-t)y_2) \leq tg(x_1,y_1) + (1-t)g(x_2,y_2)$.}, showing the potential benefits of appling network slicing to the RAN. In some cases, slicing the RAN may increase the total experienced spectral efficiency, which, in turn, translates into higher operator's revenues. Tighter conditions on the case studies when a sliced network outperforms a monolithic network structure are formulated as follows.

\begin{lemma}\label{lemma:half_sliced}
The sum-PSE of two tenants with network slicing is always greater than the sum of the PSEs of the tenants in a non-sliced network where the tenants equally split the total transmit power and available bandwidth, i.e., ${ \rm PSE_{\rm{T1}}+\rm PSE_{\rm{T2}}} > 2 {\rm {PSE}_{{NoSlicing}} }(P_{\rm{tot}}/2, B_{\rm{tot}}/2)$, with $B_{\rm T1}+B_{\rm T2} = B_{\rm{tot}}, B_{\rm T1} \neq B_{\rm T2}$ or $P_{\rm T1}+P_{\rm T2} = P_{\rm{tot}}, P_{\rm T1} \neq P_{\rm T2}$. If a uniform (equal) distribution of transmit power and bandwidth among the tenants is assumed, i.e., $\rm PSE_{\rm{T1}} = \rm PSE_{\rm{T2}}$, then ${\rm PSE_{\rm{T1}}+\rm PSE_{\rm{T2}}} = 2 {\rm {PSE_{{NoSlicing}}}}(P_{\rm tot}/2,B_{\rm tot}/2)$.
\end{lemma}
\begin{sketch}
Assuming the same MT densities, i.e., $\lambda_{\rm{T1}}=\lambda_{\rm{T2}}$ and relying on the convexity property, it yields the following:
\begin{equation}
\begin{array}{l}
{\rm {PSE_{\text{NoSlicing}}}}\!\!\left((\frac{B_{\rm{T1}}}{2}+\frac{B_{\rm{T2}}}{2}),\!(\frac{P_{\rm{T1}}}{2}+\frac{P_{\rm{T2}}}{2})\right)\!\leq\! \frac{\rm PSE_{\rm T1}}{2}\! +\! \frac{\rm PSE_{\rm T2}}{2}; \vspace{1em}\\
{\rm{ PSE_{\text{NoSlicing}}}}\left(\frac{B_{tot}}{2},\frac{P_{tot}}{2}\right) \leq \frac{\rm PSE_{\rm T1}+\rm PSE_{\rm T2}}{2}.
\end{array}
\end{equation}
If $B_{\rm T1} = B_{\rm T2}$ and $P_{\rm T1} = P_{\rm T2}$, then $B_{\rm T1} = B_{\rm {tot}}/2$, $P_{\rm T1} = P_{\rm{tot}}/2$, $\rm PSE_{\rm T1} = \rm PSE_{\rm T2}$, and we have:
\begin{equation}
\begin{array}{l}
{\rm{ PSE_{\text{NoSlicing}}}}\!\!\left(\frac{B_{\rm{tot}}}{2},\frac{P_{\rm{tot}}}{2}\right)\!\!\leq\! {\rm {PSE}}_{\rm T1}\!\!=\!\!{\rm{ PSE}}_{\rm T1}\!\!\left(\frac{B_{\rm{tot}}}{2},\frac{P_{\rm{tot}}}{2}\!\right)\!\!,
\end{array}
\end{equation}
where ${\rm {PSE}}_{\rm T1}(\frac{B_{\rm{tot}}}{2},\frac{P_{\rm{tot}}}{2}) = {\rm{ PSE}}_{\text{NoSlicing}}(\frac{B_{\rm{tot}}}{2},\frac{P_{\rm{tot}}}{2})$.
This concludes our proof. \qed
\end{sketch}
With the aid of this lemma, we can improve the system performance, by designing an admission control scheme for cellular networks that exploits network slicing and opportunistically admits subsets of tenants that maximize the PSE. This is discussed in Section~\ref{sect:gen_problem}. In particular, the following important proposition holds.
\begin{proposition}
\label{prop:alpha}
Problem~\texttt{Bi-Sharing} admits a feasible solution even if $\alpha_{\rm T1}+\alpha_{\rm T2} \geq 1$.
\end{proposition}
\begin{sketch}
The proof is obtained by combining Lemma~\ref{lemma:unsliced} and Lemma~\ref{lemma:half_sliced}. Let us consider two tenants sharing the total available bandwidth and transmit power. From Lemma~\ref{lemma:unsliced}, there is a non-negligible probability that $\rm PSE_{\rm{T1}} + \rm PSE_{\rm{T2}}= \alpha_{\rm{NoSlicing}} \rm PSE_{\rm{NoSlicing}}$, where $\alpha_{\rm{NoSlicing}}>1$. From Problem~\texttt{Bi-Sharing}, we obtain $\rm PSE_{\rm{Ti}} = \alpha_{\rm{Ti}} \rm PSE_{\rm{NoSlicing}}$, and thus, $\alpha_{\rm{T1}} + \alpha_{\rm{T2}}= \alpha_{\rm{NoSlicing}} \geq 1$. Therefore, Problem~\texttt{Bi-Sharing} admits a feasible solution for $\alpha_{\rm T1}+\alpha_{\rm T2} \geq 1$ if $B_{\rm{T1}}+B_{\rm{T2}}\leq B_{\rm{tot}}$ and $P_{\rm{T1}}+P_{\rm{T2}}\leq P_{\rm{tot}}$.
\qed
\end{sketch}
This proposition is a key-finding of this work: telecom operators can slice their radio access resources among the tenants and \emph{achieve a sum-throughput higher than that achieved without slicing the RAN}, i.e., by sharing the available resources among the tenants without performance guarantees. In the next section, we discuss the solution of Problem~\texttt{Bi-Sharing}.

\subsection{Lagrange Decomposition}
\label{sect:decom}
In order to solve Problem~\texttt{Bi-Sharing}, we propose to apply the Lagrange duality theorem~\cite{Boyd2004}. Let us define the Lagrangian $\mathcal{L} : R^m\times R^n \to R$, where $m=6$ is the number of decision variables and $n=2$ is the number of constraints, as follows:
\begin{equation} 
\label{eq:duality}
\begin{array}{l}
\mathcal{L}(\mu_{\rm{T1}},\mu_{\rm{T2}},B_{\rm T1},P_{\rm T1},B_{\rm T2},P_{\rm T2}) = \\
1 - \mu_{\rm{T1}}\left(\alpha_1 {\rm{ PSE_{\text{NoSlicing}}}} - K_1 B_{\rm T1} (1-e^{-\frac{P_{\rm T1}}{B_{\rm T1}}^{(2/\beta)} k_2}) \right) - \\
\mu_{\rm{T2}}\left(\alpha_2 {\rm{ PSE_{\text{NoSlicing}}}} - K_1 B_{\rm T2} (1-e^{-\frac{P_{\rm T2}}{B_{\rm T2}}^{(2/\beta)} k_2}) \right).
\end{array}
\end{equation}
We can derive the Lagrange dual function $g(\mu_{\rm{T1}},\mu_{\rm{T2}}) = \inf\limits_{B_{\rm T1},B_{\rm T2},P_{\rm T1},P_{\rm T2}} \mathcal{L}(\mu_{\rm{T1}},\mu_{\rm{T2}},B_{\rm T1},P_{\rm T1},B_{\rm T2},P_{\rm T2})$ that satisfies the constraints $B_{\rm T1} + B_{\rm T2} \leq B_{\rm{tot}}$ and $P_{\rm T1} + P_{\rm T2} \leq P_{\rm{tot}}$. The unconstrained dual problem can be formulated as follows.

\vspace{2mm}\noindent \textbf{Problem}~\texttt{Bi-Sharing(DUAL)}:
\begin{equation*}
\label{pr:bi-sharing-dual}
\begin{array}{ll}
\text{maximize}  & g(\mu_{\rm{T1}},\mu_{\rm{T2}})\\
\text{subject to } & \mu_{\rm{T1}},\mu_{\rm{T2}} \geq 0.
\end{array}
\end{equation*}

This problem can be solved by using the iterative sub-gradient update method to optimize the Lagrange multipliers $\mu_{\rm{T1}}$ and $\mu_{\rm{T2}}$:
\begin{equation}
\label{eq:update1}
\begin{array}{ll}
 &\mu_{\rm{T1}}^{(k+1)}\!\! =\!\! \left [ \mu_{\rm{T1}}^{(k)}\!\! + \zeta^{(k)} \Big ( \frac{\partial \mathcal{L}(\mu_{\rm{T1}},\mu_{\rm{T2}},B_{\rm T1},P_{\rm T1},B_{\rm T2},P_{\rm T2})^{(k)}}{\partial \mu_{\rm{T1}}}  \Big ) \right ]^+\\
 &\mu_{\rm{T2}}^{(k+1)}\!\! =\!\! \left [ \mu_{\rm{T2}}^{(k)}\!\! + \zeta^{(k)} \Big ( \frac{\partial \mathcal{L}(\mu_{\rm{T1}},\mu_{\rm{T2}},B_{\rm T1},P_{\rm T1},B_{\rm T2},P_{\rm T2})^{(k)}}{\partial \mu_{\rm{T2}}} \Big ) \right ]^+
\end{array}
\end{equation}
where $\zeta^{(k)}$ is defined as the step-size and can be chosen as follows (\cite{SHI2005}):
\begin{equation}
\label{eq:zeta_lagr}
\zeta^{(k)} = \frac{\nu^{(k)}}{\norm{\frac{\partial \mathcal{L}(\mu_{\rm{T1}},\mu_{\rm{T2}},B_{\rm T1},P_{\rm T1},B_{\rm T2},P_{\rm T2})^{(k)}}{\partial \vv{\mu}}}_2}
\end{equation}
where $\nu^{(k)}=\norm{\vv{\mu}^{(k)}-\vv{\mu}^{(k-1)}}_2$ with $\norm{\cdot}_2$ denoting the norm-2 operation whereas $\zeta^{(k)}$ must be greater than zero. The iterative process stops when the convergence is reached and the optimal Lagrange multipliers $\mu_{\rm{T1}}^{(*)},\mu_{\rm{T2}}^{(*)}$ are found, i.e., when $\mu_{\rm{T1}}^{(k+1)} = \mu_{\rm{T1}}^{(k)}$ and $\mu_{\rm{T2}}^{(k+1)} = \mu_{\rm{T2}}^{(k)}$. 

If two tenants are considered, the Lagrange multipliers can be obtained without using this recursive approach. This can be done as follows. Let us generalize the integrity constraint as $B_{\rm T1} + B_{\rm T2} = B_{\rm{tot}}$ and $P_{\rm T1} + P_{\rm T2} = P_{\rm{tot}}$. This allows us to derive $\mathcal{L}(\mu_{\rm{T1}},\mu_{\rm{T2}},B_{\rm T1},P_{\rm T1})$, where $B_{\rm T2}$ and $P_{\rm T2}$ can be obtained from the equalities $B_{\rm T1} + B_{\rm T2} = B_{\rm{tot}}$ and $P_{\rm T1} + P_{\rm T2} = P_{\rm{tot}}$. Therefore, we can rewrite Eq.~\eqref{eq:duality} and obtain the optimal $\mu_{\rm{T1}}^{(*)}$ and $\mu_{\rm{T2}}^{(*)}$ for given $B_{\rm T1}$ and $P_{\rm T1}$. The following is obtained:
\begin{equation}
\label{eq:update1}
\begin{array}{ll}
 &\mu_{\rm{T1}}^{(*)} = \alpha_{\rm{T1}}^{-1}\\
 &\mu_{\rm{T2}}^{(*)} = \alpha_{\rm{T2}}^{-1}.
\end{array}
\end{equation}

Based on this finding, Problem~\texttt{Bi-Sharing} can be reduced to an unconstrained feasibility problem as follows. 

\vspace{2mm}\noindent \textbf{Problem}~\texttt{Bi-Sharing(UNCONSTRAINED)}:
\begin{equation*}
\label{pr:bi-sharing-unconstr}
\begin{array}{ll}
\text{minimize}  & 1\!-\! 2 {\rm{ PSE_{\text{NoSlicing}}}}\!-\!\frac{k_1^{(\rm{T1})}}{\alpha_{\rm{T1}}}\! B_{\rm T1}\!\!\left(\! 1-e\!\!\!^{-\frac{P_{\rm T1}}{B_{\rm T1}}^{(2/\beta)} \!k_2^{(\rm{T1})}}\!\!\right) \\
				 & - \frac{k_1^{(\rm{T2})}}{\alpha_{\rm{T2}}} \left(B_{\rm{tot}}-\!B_{\rm T1}\right)\!\! \left(\!\!1-e^{-\frac{P_{\rm{tot}}-P_{\rm T1}}{B_{\rm{tot}}-B_{\rm T1}}^{(2/\beta)} \!k_2^{(\rm{T2})}}\right) \\
\text{subject to } & 0 \leq B_{\rm T1} \leq B_{\rm{tot}}; \\
				   & 0 \leq P_{\rm T1} \leq P_{\rm{tot}}.
\end{array}
\end{equation*}
Based on this result, Problem~\texttt{Bi-Sharing} can be solved with the aid of conventional numerical methods.

In the next section, we provide a generalized formulation of the problem for a multi-tenant system where $\sum_i \alpha_i \geq 1$.

\subsection{Generalized Problem Formulation}
\label{sect:gen_problem}
Let us consider a set of tenants $i\in\mathcal{I}$ each of them requesting a network slice. The objective it to efficiently split the resources of the RAN among them. This encompasses the allocation of adequate transmit powers and spectrum bandwidths to each tenant, i.e., $\vv{p}=\{P_{\rm{Ti}}\}$ and $\vv{b}=\{B_{\rm{Ti}}\}$. Based on Problem~\texttt{Bi-Sharing}, we can formulate a general optimization problem as follows.

\vspace{2mm}\noindent \textbf{Problem}~\texttt{MultiTenant-Optimizer}:
\begin{equation*}
\label{pr:mt_optimizer}
\begin{array}{ll}
\text{minimize}  & \mathbbm{1}\\
\text{subject to } & \rm{PSE}(\vv{b},\vv{p},\vv{\lambda}) \geq \vv{\alpha}\,{\rm{PS}}{{\rm{E}}_{{\rm{NoSlicing}}}} \\
				   & \norm{\vv{b}}_1 \leq B_{\rm{tot}};\\
				   & \norm{\vv{p}}_1 \leq P_{\rm{tot}}.
\end{array}
\end{equation*}
where $\norm{\cdot}_1$ is the norm-1 operator, $\vv{\lambda}=\{\lambda_{\rm{Ti}}\}$ is the set of user densities of the tenants and $\vv{\alpha}$ is the vector of SLA requirements of the tenants, which are formulated in terms of percentages of the PSE without performing network slicing, i.e., ${\rm{PS}}{{\rm{E}}_{{\rm{NoSlicing}}}}$. We can formulate the dual problem of Problem~\texttt{MultiTenant-Optimizer} and calculate the optimal Lagrange multipliers $\vv{\mu}=\{\mu_{\rm{Ti}}\}$ by using the iterative function as follows:
\begin{equation}
\label{eq:lagr_multi}
\mu_{\rm{Ti}}^{(k+1)} = \left[ \mu_{\rm{Ti}}^{(k)} +\zeta^{(k)}\left(\frac{\partial\mathcal{L}(\vv{\mu},\vv{b},\vv{p})}{\partial\mu_{\rm{Ti}}}\right)\right]^+.
\end{equation}
This leads to a problem similar to Problem~\texttt{MultiTenant-Optimizer (UNCONSTRAINED)} studied in Section~\ref{sect:decom}.

If the solution of the optimization problem consists of small values of transmit power or bandwidth, i.e., $P_{\rm{Ti}}\approx 0$ or $B_{\rm{Ti}}\approx 0$, the system discards the request of the slice that originates from tenant $i$: this is the essence of the proposed admission control protocol. To speed up the admission control phase, we propose to pre-filter the requests of RAN slices beforehand. In particular, we propose to process all the slices that need less than half of the non-sliced spectral efficiency, i.e., the slices with $\alpha_{\rm{Ti}}<= 0.5$ are processed for optimal resource allocation while the others are discarded. This approach is motivated by the finding in Lemma~\ref{lemma:half_sliced}. This leads the sliced RAN to have a sum-throughput that is in general close and may be higher than its non-sliced counterpart, as shown in Sec.~\ref{sect:pe_slicing_benefit}. More precisely, our proposed admission control scheme works as follows: if, after the first assignment round, the sum-throughput of the admitted tenants is below the non-sliced PSE, then other slice requests from other tenants are considered in ascending order until the best sum-throughput is found. This is the essence of STORNS.

In Alg.~\ref{algo:qlea}, we provide the pseudo-code description of STORNS, which yields the set bandwidths and transmit powers for each admitted slice. Steps $3-7$ are repeated until the Lagrange multipliers reach convergence. The speed and accuracy of the proposed algorithm are determined by the step-size $\zeta^{(k)}$, defined in Eq.~\eqref{eq:zeta_lagr}. An empirical analysis of the convergence of the algorithm is provided in Sec.~\ref{sec:pe_complex}.

\begin{algorithm}[t]
\centering
\vspace*{-1pt}
\begin{framed}

\begin{enumerate}
\small
\vspace*{-5pt}
\item Initialise set $\vv{\mu}$ and $\gamma$.
\item Initialise sets $\vv{b} \leftarrow 0$, $\vv{p} \leftarrow 0$ and value $k\leftarrow 0$.
\item Solve Problem~\texttt{MultiTenant-Optimizer (DUAL)} (with INPUT $\vv{b}^{(k)}$ and $\vv{p}^{(k)}$) and get $\vv{\mu}^{(k)}$.
\item Calculate $\vv{\mu}^{(k+1)}$ based on Eq.~\eqref{eq:lagr_multi}.
\item Update $\vv{\zeta}^{(k+1)}$ based on Eq.~\eqref{eq:zeta_lagr}.
\item Solve Problem~\texttt{MultiTenant-Optimizer (UNCONSTRAINED)} (with INPUT $\vv{\mu}^{(k+1)}$) and get $\vv{b}^{(k+1)}$ and $\vv{p}^{(k+1)}$.
\item If ($\vv{\mu}^{(k+1)} \neq \vv{\mu}^{(k)}$), then increase $k=k+1$ and \underline{Go to step (3)}.
\item Mark $\vv{\mu}^{(*)} = \vv{\mu}^{(k+1)}$ as the optimal Lagrange multipliers.
\item Solve Problem~\texttt{MultiTenant-Optimizer (UNCONSTRAINED)} (with INPUT $\vv{\mu}^{(*)}$) and get the optimal solution of $\vv{b}^{(*)}$ and $\vv{p}^{(*)}$.
\end{enumerate}
\vspace*{-10pt}
\end{framed}
\caption{Stochastic RAN Slicer (STORNS)}
\label{algo:qlea}
\vspace*{-8.5pt}
\end{algorithm}

\section{Performance evaluation}
\label{s:results}

We have carried out an extensive simulation study to $i$) validate the analytical framework based on stochastic geometry theory, $ii$) show the near-optimality of the proposed low-complex algorithm (STORNS) and $iii$) prove the finding that slicing the RAN may provide a higher sum-PSE compared to its non-sliced counterpart. We have implemented frameworks and algorithms by using commercial mathematical tools, such as MATLAB and MATHEMATICA. We consider different random instances of cellular network deployments based on the PPP model, as explained in Sec.~\ref{PPP_CellularModeling}. Unless otherwise stated, the simulations parameters are those reported in Table~\ref{tab:system_param}, summarizing the Urban Micro-cell model (UMi) which is in agreement with IMT ITU-R specifications~\cite{imt_itur}. The pathloss exponent is chosen based on the empirical evaluations performed in~\cite{pathloss_rappaport2016}.

\begin{table}[h!]
\caption{Simulation parameters (ITU UMi~\cite{imt_itur})}
\label{tab:system_param}
\scriptsize
\centering
\begin{tabular}{|c|c|}
\hline
\textbf{System Parameters} & \textbf{Values}\\
\hline
Inter-site distance (ISD) & $200$ m\\
Base station density ($\lambda_{BS}$) & $ (\pi\,\text{ISD}^2)^{-1} $\\
Mobile terminal density ($\lambda_{T}$) & $ 100\,\lambda_{BS}$\\
Decoding threshold ($\gamma_I$) & $0$ dB\\
Detecting threshold ($\gamma_A$) & $0$ dB\\
Carrier frequency ($f_c$) & $2.1\cdot 10^9$ Hz\\
Transmission wavelength ($\lambda$) & $\frac{3\cdot 10^8}{f_c}$ m\\
Path-loss constant ($\kappa$) & $(\frac{4\pi}{\lambda})^2$\\
Noise power spectral density $(N_0)$ & $-174$ dBm/Hz\\
($\tau_A$) & $(\kappa\,\gamma_A\,N_0)^{-1}$\\
Path-loss exponent ($\beta$)~\cite{pathloss_rappaport2016} & $3.5$\\
Bandwidth ($B_{\rm{tot}}$) & $20\cdot 10^6$ Hz\\
Transmit power ($P_{\rm{tot}}$) & $43$ dBm\\
Slice generation process (mean $\mu_{\text{distr}}$) & $10$\\
Slice generation process (variance $\sigma^2_{\text{distr}}$) & $5$\\
Simulation instances (STORNS) & $1000$ \\
\hline
\end{tabular}
\end{table}

\subsection{Stochastic Framework Validation}
\begin{figure}[t!]
      \centering
      \includegraphics[trim = 0mm 0mm 0mm 0mm, clip, width=0.8\linewidth ]{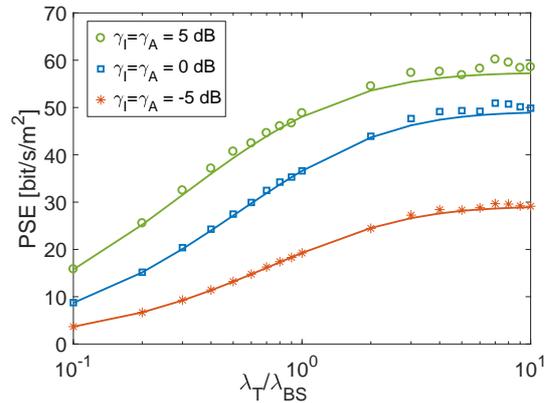}
      \caption{Framework validation. Solid lines: Eq.~\eqref{Eq_6}. Markers: Monte Carlo simulations.}
      \label{fig:validation}
\end{figure}

In Fig.~\ref{fig:validation}, the PSE in Eq.~\eqref{Eq_6} is validated against Monte Carlo simulations. The simulations are obtained by considering several instances for the locations of BSs and MTs, which follow two independent PPPs. For each network realization, the potential throughput of each MT is computed based on its definition in Eq.~\eqref{Eq_2}. It is worth mentioning that, to make the validation sound, none of the mathematical equations in Sec.~\ref{s:system} are used. In particular, the number of users per cell is directly obtained from Monte Carlo simulations. The network spectral efficiency is obtained by summing the potential throughput of all the MTs and normalizing it to the area of the network. Fig.~\ref{fig:validation} shows a good agreement between analysis and simulations. The small inaccuracies for a large ratio of the density of MTs and BSs is due to the limited number of network realizations that can be simulated in a reasonable amount of time. We note, in particular, that PSE increases as a function of the ratio of densities of MTs and BSs, but saturates as this ratio gets large. This is due to the fact that, as the number of MTs increases, all the BSs are activated and the bandwidth allocated to each MT decreases at the same rate as the number of MTs per unit area. Mathematically speaking, the $L(\cdot)$ function in Eq.~\eqref{Eq_6} tends to one as the ratio $\lambda_T/\lambda_{\rm{BS}}$ increases towards infinity.

\subsection{Optimality of STORNS}
\label{sec:pe_optm}
In this section, we validate our algorithm, STORNS, against a benchmark optimal algorithm that is obtained by using a brute-force optimization method that is denoted by \emph{OPT}. In particular, OPT is obtained by means of an exhaustive greedy search algorithm that explores all possible solutions of Problem~\texttt{MultiTenant-Optimizer}. Due to complexity issues, we are able to employ this method for up to $6$ tenants.

\begin{figure}[t!]
      \centering
      \includegraphics[trim = 0mm 0mm 0mm 0mm, clip, width=0.8\linewidth ]{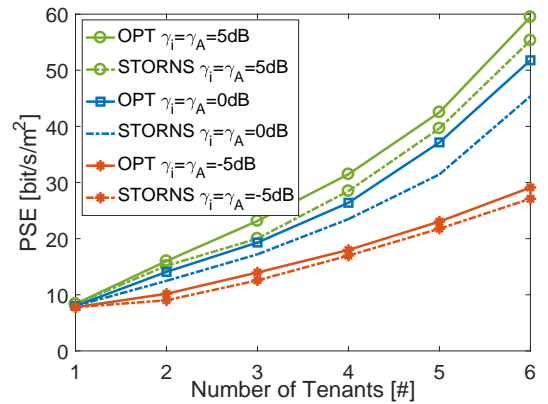}
      \caption{Optimality study of STORNS}
      \label{fig:optimality}
\end{figure}

In Fig.~\ref{fig:optimality}, we show numerical results by setting $\lambda_{\rm{T}}=100\lambda_{\rm{BS}}$ and by considering different thresholds $\gamma_I$ and $\gamma_A$. The objective is to compare STORNS and OPT as the number of tenants requesting a RAN slice increases. We assume that each additional tenant asks for a fraction $\alpha_{\rm{Ti}}$ of $\rm PSE_{\rm{NoSlicing}}$ that is drawn from a normal distribution with mean $\mu_{\text{distr}}$ and variance $\sigma_{\text{distr}}$, as defined in Table~\ref{tab:system_param}.

The larger the number of slice requests, the higher the PSE. We observe that STORNS exhibits near-optimal performance and the gap with respect to OPT is around $6\%$, $8\%$ and $12\%$ for $\gamma_I=\gamma_A=-5$dB,$0$dB and $5$dB, respectively.

\subsection{The RAN slicing benefits}
\label{sect:pe_slicing_benefit}
\begin{figure}[t!]
      \centering
      \includegraphics[trim = 0mm 0mm 0mm 0mm, clip, width=0.8\linewidth ]{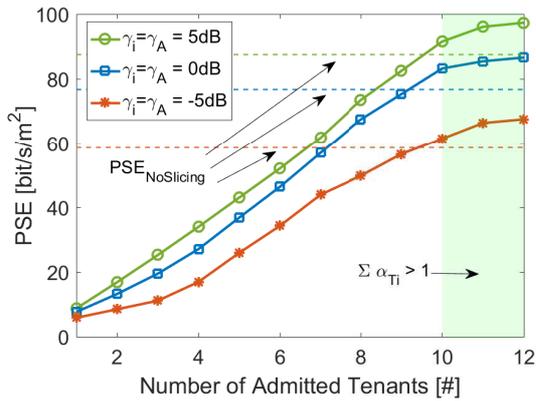}
      \caption{Potential benefits of network slicing}
      \label{fig:throughput}
\end{figure}

In Fig.~\ref{fig:throughput}, we illustrate the performance offered by STORNS as a function of the number of tenants admitted into the cellular network. More precisely, we provide numerical evidence that STORNS is capable of appropriately admitting tenants and allocating their slices (i.e., bandwidth and power) in a way that the PSE of the sliced RAN is higher than its monolithic network counterpart that does not exploit network slicing. We assume a network slices demand up to $32$ network slice requests. In addition, we assume that each tenant requests, on average, a network slice that provides a PSE that is 10$\%$ of the achievable PSE without using slicing. This implies that the non-sliced network would be able to admit, on average, up to ten tenants (non-shaded region in the figure)\footnote{It is worth nothing that the overall PSE of $10$ admitted network slice requests that corresponds to $100\%$ of PSE$_\text{NoSlicing}$ is slightly above the dashed line shown in the figure.
This is due to the randomness of the network slice generation process. In fact, 10$\%$ is only the mean value (see Table~\ref{tab:system_param}).}. By using STORNS, we can accommodate a larger number of tenants and achieve a sum-PSE that is higher than the non-sliced sum-PSE (shaded region in the figure). This is possible by admitting the ``best'' network slice requests among the $32$ available and by optimally allocating the transmit power and bandwidth to each of them. STORNS allows telecom operators to achieve up to $120\%$ of the throughput of a monolithic non-sliced cellular network. This motivates telecom operators to use network slicing not only as a means for accommodating the specific request of vertical industries, but also as a powerful means for enhancing the overall network performance and, in turn, for increasing their revenues by simply sharing their physical infrastructure among multiple tenants.

In Table~\ref{tab:max_perf}, we evaluate the gain provided by RAN slicing by using STORNS. The gain is defined as $\rm{PSE}_{\text{max}}/\rm{PSE}_{\text{NoSlicing}}$, where $\rm{PSE}_{\text{max}} = \sum_{i} \rm{PSE}_{\rm{Ti}}, \forall i$ admitted. We evaluate different user densities $\lambda_{T}$ and threshold parameters $\gamma_I,\gamma_A$. When the average number of users increases, we note that network slicing provides additional performance gains (about $19\%$).
\vspace{-2.5mm}

\begin{table}[h!]
\caption{RAN Slicing Gain}
\label{tab:max_perf}
\centering
\begin{tabular}{|c|c|c|c|}
\hline
 & $\frac{\lambda_T}{\lambda_{\rm{BS}}} = 50$ & $\frac{\lambda_T}{\lambda_{\rm{BS}}} = 200$ & $\frac{\lambda_T}{\lambda_{\rm{BS}}} = 500$\\
\hline
$\gamma_I=\gamma_A=5$dB & $7.17\%$ &$13.2\%$ &$18.1\%$\\
$\gamma_I=\gamma_A=0$dB & $6.63\%$ & $16.6\%$ &$18.8\%$\\
$\gamma_I=\gamma_A=-5$dB & $4.52\%$ &$14.8\%$ &$17.69\%$\\
\hline
\end{tabular}
\vspace{-1.5em}
\end{table}

\subsection{Algorithm complexity}
\label{sec:pe_complex}
\begin{figure}[t!]
      \centering
      \includegraphics[trim = 0mm 0mm 0mm 0mm, clip, width=0.8\linewidth ]{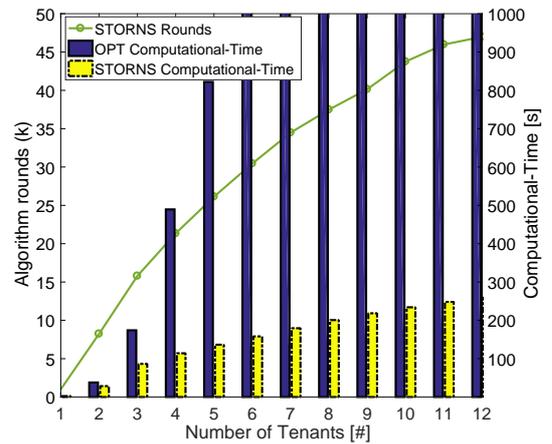}
      \caption{Computational Analysis}
      \label{fig:computation}
\end{figure}

We study the complexity of our algorithm against that of the exhaustive greedy search. The main parameter for STORNS is the number of rounds to converge and to compute the optimal Lagrange multipliers, as explained in Sec.~\ref{sect:gen_problem}. In Fig.~\ref{fig:computation}, we show with a solid green line the number of rounds ($k$) that are needed to converge while increasing the number of tenants requesting a RAN slice. The behavior of the curve unveils that the complexity of our algorithm does not exponentially increases with the number of constraints (i.e., the number of tenants) of the optimization problem (Problem~\texttt{MultiTenant-Optimizer}) but it converges to a stable number of iterations. On the right y-axis of Fig.~\ref{fig:computation}, we compare the computational time for solving the optimization problem and compare STORNS against OPT. We evince that STORNS is capable of achieving near-optimal performance with a limited complexity compared to greedy approaches.

\vspace{3mm}
\section{Conclusions}
\label{s:concl}
We have analyzed the benefits of applying network slicing to radio access networks. In particular, we have considered network slice requests with diverse service level agreements (SLAs) in terms of required average throughput per tenant. To analytically formulate the problem, we have capitalized on stochastic geometry theory, which allowed us to consider cellular network topologies in a tractable yet sufficiently realistic manner. We have introduced a new mathematical formulation for network slicing throughput and have defined an optimization problem to design an admission control, \emph{STORNS},  that identifies the best tenants to be admitted into the network along with their spectrum and transmit power allocation such that the overall system throughput is maximized. 

Our results have shown through mathematical proofs, numerical and simulation results that networks where network slicing is applied can achieve a higher network throughput than non-sliced ones. Finally, we have provided quantitative results of reduced computational complexity of STORNS as compared to brute-force optimization methods. Our work puts forth network slicing as a suitable approach for optimizing the radio resource utilization of future sliced cellular networks. 
\vspace{3mm}


\bibliographystyle{IEEEtran}
\bibliography{bibliography}


\end{document}